\documentclass[runningheads]{llncs}
\usepackage[T1]{fontenc}
\usepackage{graphicx}
\usepackage{marvosym}
\usepackage{xcolor}
\usepackage{multirow}
\definecolor{rqbg}{rgb}{0.95,0.95,0.95}
\usepackage{hyperref}
\usepackage{booktabs}      
\usepackage{array}         
\usepackage{xcolor}        
\usepackage{colortbl}      
\usepackage{xurl} 
\usepackage{amsmath} 

\makeatletter
\renewcommand{\paragraph}{%
  \@startsection{paragraph}{4}%
  {\z@}{0.2ex \@plus 1ex \@minus .2ex}{-1em}%
  {\normalfont\normalsize\bfseries}%
}

\makeatother
\newcommand{\Paragraph}[1]{\paragraph{{#1.}}}

\usepackage{tcolorbox}
\usepackage{float}
\tcbuselibrary{skins,breakable}

\begin{document}
%





\title{Can Multimodal LLMs ‘See’ Science Instruction? Benchmarking Pedagogical Reasoning in K--12 Classroom Videos}

%
\titlerunning{Benchmarking Pedagogical Reasoning in K--12 Science Videos}
%
%
\author{Yixuan Shen\inst{1}\thanks{Equal contribution.} \and
Peng He\inst{2}\textsuperscript{$\star$}\and 
Honglu Liu\inst{2,3} \and
Jinxuan Fan\inst{1} \and
Yuyang Ji\inst{1} \and \\
Tingting Li\inst{2} \and 
Tianlong Chen\inst{4} \and
Kaidi Xu\inst{5} \and
Feng Liu\inst{1}\textsuperscript{\Letter}
}
\authorrunning{Y. Shen, P. He et al.}
%
\institute{Department of Computer Science, Drexel University 
\and
Department of Teaching and Learning, Washington State University
\and
College of Chemistry, Beijing Normal University
\and
Department of Computer Science, University of North Carolina at Chapel Hill
\and
Department of Data Science, City University of Hong Kong
}
\maketitle              
\begingroup
\renewcommand{\thefootnote}{} 
\footnotetext{\Letter\ Corresponding author. \texttt{fl397@drexel.edu}}
\endgroup

\begin{abstract}
K--12 science classrooms are rich sites of inquiry where students coordinate phenomena, evidence, and explanatory models through discourse; yet, the multimodal complexity of these interactions has made automated analysis elusive.
Existing benchmarks for classroom discourse focus primarily on mathematics and rely solely on transcripts, overlooking the visual artifacts and model-based reasoning emphasized by the Next Generation Science Standards (NGSS).
We address this gap with \textbf{SciIBI}, the first video benchmark for analyzing science classroom discourse, featuring 113 NGSS-aligned clips annotated with Core Instructional Practices (CIP) and sophistication levels.
By evaluating eight state-of-the-art LLMs and Multimodal LLMs, we reveal fundamental limitations: current models struggle to distinguish pedagogically similar practices, suggesting that CIP coding requires instructional reasoning beyond surface pattern matching. Furthermore, adding video input yields inconsistent gains across architectures.
Crucially, our evidence-based evaluation reveals that models often succeed through surface shortcuts rather than genuine pedagogical understanding.
These findings establish science classroom discourse as a challenging frontier for multimodal AI and point toward human-AI collaboration, where models retrieve evidence to accelerate expert review rather than replace it.
\href{https://vilab-group.com/project/sciibi}{Project} 

\end{abstract}
%

\section{Introduction}
Science education reforms increasingly view classroom interaction as a learning community in which teachers mediate students' participation in disciplinary practices, rather than simply delivering static knowledge for memorization~\cite{he2023transforming}.
While the National Research Council~\cite{nrc2012framework} and the Next Generation Science Standards (NGSS)~\cite{ngss2013} emphasize that teachers should guide students to learn science by integrating disciplinary core ideas, scientific practices, and crosscutting concepts, teachers often struggle to visualize how these abstract standards translate into ideal instruction moment-to-moment. This ambiguity creates significant challenges for teacher preparation and professional learning at scale~\cite{sherin2009effects}.

A long-standing response in the learning sciences is the use of classroom video and transcripts for practice-based instructional analysis~\cite{goldman2007video}.
Such analyses provide a window into how teachers elicit student thinking, organize inquiry activities, and press for evidence-based explanations---insights that directly inform teacher professional development.
However, high-quality coding typically requires trained human raters applying detailed rubrics, making annotation time-consuming and costly, limiting routine use in authentic school settings~\cite{worsley2018mla}.

\begin{figure}[t]
\includegraphics[width=1\textwidth]{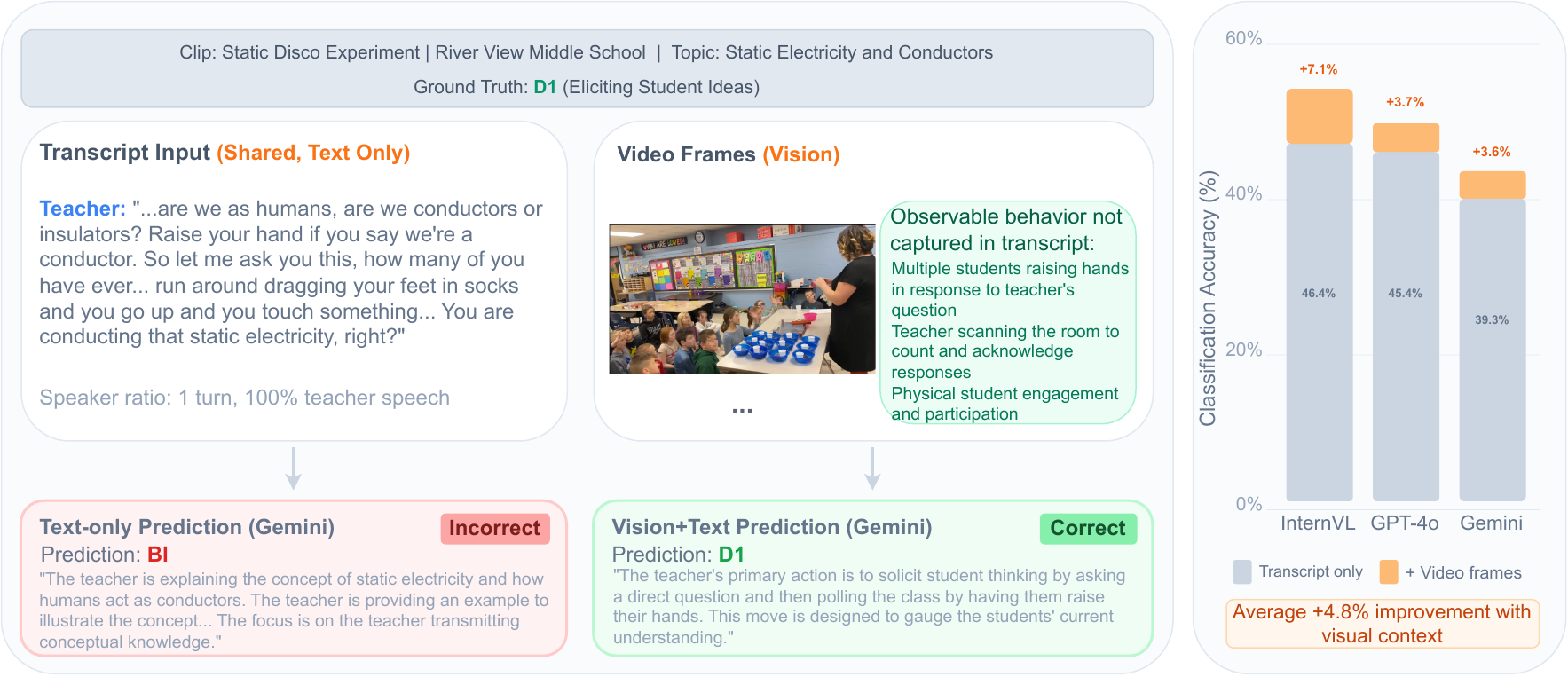}
\vspace{-5mm}
\caption{\textbf{The impact of visual context on instructional practice coding.} \textit{Left:} Text-only models fail to detect student engagement (\emph{e.g.}, raising hands) from the transcript alone, misclassifying the clip as a lecture (Big Idea). \textit{Right:} Multimodal models (Vision+Text) utilize visual cues to correctly identify "Eliciting Student Ideas" (D1), yielding an average accuracy improvement of 4.8\% across evaluated MLLMs.}
\label{fig:fig1}
\vspace{-3mm}
\end{figure}

Most benchmarked work on the automated analysis of classroom discourse is math-focused and transcript-based.
The TalkMoves dataset (K--12 math lesson transcripts)~\cite{suresh2022talkmoves} remains a dominant benchmark; fine-tuned transformers achieve F1 scores around 79\%~\cite{suresh2022fine}, and recent LLM-based approaches show competitive results~\cite{moreau2024classifying}.
Subsequent studies have explored explainability~\cite{wang2023aied_xai} and zero-shot comparisons~\cite{wang2023chatgpt}.
While multimodal approaches have been explored~\cite{hou2025multimodal}, evaluations remain limited to mathematics, and evidence regarding the benefits of video beyond transcripts remains mixed.

Science classrooms present a stronger case for multimodal analysis than mathematics, as science teaching and learning inherently require multimodal engagement, such as conducting lab activities or making sense of phenomena through drawn and written responses.
Consequently, Core Instructional Practice (CIP) coding often depends on the interplay among phenomena, evidence, and models unfolding in the classroom: student drawings on whiteboards, data tables, manipulated materials, and teacher gestures toward shared representations.
These ``public inscriptions''\cite{latour1986visualization} carry pedagogical intent that transcripts alone cannot fully capture (see Fig.\ref{fig:fig1}).
Despite this, science classroom discourse coding is notably under-studied; to our knowledge, no existing benchmark systematically evaluates automated CIP coding for science instruction, leaving open the question of whether findings from mathematics transfer to contexts with different interaction patterns and artifact usage.
Research in multimodal learning analytics similarly indicates that educational signals are distributed across speech, action, and artifacts~\cite{ochoa2016mla,worsley2018mla}.
For example, distinguishing sensemaking around material activity from evidence-pressing moves may require observing \emph{what} artifact the teacher references and \emph{how} students interact with it.
Multimodal LLMs (MLLMs), which process text with visual inputs~\cite{openai2023gpt4,team2023gemini,liu2023llava}, offer a promising path to capture this richer context, yet their readiness for science classroom analytics remains untested.

Beyond classification accuracy, deployment in teacher-facing settings demands \emph{trustworthiness}.
A model that outputs labels without grounded support is less useful than one that provides verifiable justification, as the latter enables efficient human verification and supports teacher reflection~\cite{jacobs2022promoting,holstein2019co}.
Traditional evaluations report aggregate F1 scores but offer limited insight into whether correct predictions are made for the right reasons; models may succeed through surface cues rather than genuine recognition of pedagogical intent.
We therefore introduce an \emph{evidence-based evaluation} protocol: models must produce supporting evidence alongside predictions---specifically, text evidence (quoted spans or sentence indices) and/or temporal evidence (timestamps or clip intervals)---and we assess whether that evidence is sufficient and aligned with the predicted category.



In this paper, we frame science classroom video analysis as a \emph{Core Instructional Practice} coding task grounded in Inquiry-Based Instruction (IBI).
Windschitl \emph{et al.}\cite{windschitl2012proposing} proposed the Core Instructional Practices (CIP) framework to support teachers designing and organizing classroom discourse and activities aligned with educational standards.
These practices align with NGSS goals of linking observable phenomena with underlying mechanisms through iteratively refined models and evidence\cite{nrc2012framework,ngss2013}.
These considerations motivate an evaluation setting reflecting practical deployment constraints, where labeled science data are scarce and fine-tuning is often infeasible.
We examine three prompting settings common in foundation-model research: zero-shot, few-shot~\cite{brown2020language}, and chain-of-thought~\cite{wei2022chain,kojima2022large}.
Our work is guided by four research questions:

\vspace{-3mm}
\begin{center}
	\setlength\fboxrule{0.0pt}
	\noindent\fcolorbox{black}[rgb]{0.95,0.95,0.95}{\begin{minipage}{1\columnwidth}
(\textbf{Q1}) How do state-of-the-art MLLMs perform on CIP coding, and what does this reveal about their pedagogical reasoning capabilities?

\vspace*{0.1cm}
(\textbf{Q2}) When does visual input provide value beyond transcripts, and does this vary across different CIP categories?

\vspace*{0.1cm}
(\textbf{Q3}) 
How do classification accuracy and evidence quality relate?

\vspace*{0.1cm}
(\textbf{Q4}) What failure modes emerge, and how should they inform human--AI deployment workflows?
	\end{minipage}}
\end{center}

In light of this, we introduce \textbf{SciIBI} (\textbf{Sci}ence \textbf{I}nquiry-\textbf{B}ased \textbf{I}nstruction), a benchmark of K--12 science classroom video clips sourced from YouTube and annotated using the CIP framework.
SciIBI enables controlled comparisons across input modalities (text-only vs.\ vision+text) and prompting settings (zero-shot, few-shot, chain-of-thought), and supports evidence-based diagnostics by requiring models to localize supporting evidence for their predictions.

In summary, the contributions of this work include:

$\diamond$ We collect \textbf{SciIBI}, the first (to our knowledge) K--12 science classroom video benchmark for CIP coding, consisting of 113 clips with an accompanying sophistication probe and a reproducible evaluation protocol.

$\diamond$ We conduct controlled comparisons across input modalities (text-only vs.\ vision+text) and prompting settings, identifying specifically when visual context aids performance and when it does not. Together, these contributions establish SciIBI as a diagnostic tool for understanding model limitations rather than merely a leaderboard for optimizing classification accuracy.

$\diamond$ We introduce an evidence-aligned scoring protocol requiring models to output supporting evidence alongside predictions. Through expert assessment of evidence sufficiency, alignment, and completeness, we reveal critical cases where accuracy and evidence quality diverge.
\section{Related Work}

\Paragraph{Transcript-Based Classroom Discourse Coding}
Analyzing classroom discourse has a long history in education research.
Early frameworks such as Initiation--Response--Evaluation (IRE)~\cite{mehan1979learning} and subsequent shifts toward dialogic instruction~\cite{cazden2001classroom,michaels2008deliberative} established discourse coding as a primary lens for understanding how teachers support student reasoning.
Accountable Talk~\cite{michaels2008deliberative} and related schemes operationalize \emph{talk moves}---teacher utterances that press for reasoning, revoice student ideas, or promote peer engagement---as indicators of instructional quality.
Computational approaches emerged with the availability of large-scale transcript corpora.
The TalkMoves dataset~\cite{suresh2022talkmoves}, comprising 567 K--12 mathematics lesson transcripts with turn-level annotations, has become the \emph{de facto} benchmark.
Suresh \emph{et al}.~\cite{suresh2022fine} fine-tuned transformer models on this data, and later work scaled coding using large language models~\cite{moreau2024classifying}.
Subsequent studies have examined explainability~\cite{wang2023aied_xai}, zero-shot LLM performance~\cite{wang2023chatgpt}, and specific discourse phenomena such as conversational uptake~\cite{demszky2021ncte}.
Despite this progress, existing benchmarks predominantly target mathematics and rely solely on transcripts.
Science classrooms differ in fundamental ways: Model-Based Inquiry (MBI) emphasizes coordinating phenomena, evidence, and models through artifact-mediated sensemaking~\cite{windschitl2012proposing}, a process that may not transfer directly from math-focused corpora.
To our knowledge, there is no widely used benchmark that evaluates automated CIP coding in science instruction.

\Paragraph{Multimodal Classroom Analytics from Video}
Video has long been central for capturing the complexity of classroom interaction.
Goldman \emph{et al}.~\cite{goldman2007video} established video analysis as a core methodology in the learning sciences, and Sherin and van Es~\cite{sherin2009effects} demonstrated its value for developing teachers' professional vision.
However, these approaches rely on expert human viewing and coding, which scales poorly.
Multimodal Learning Analytics (MLA) aims to automate this analysis using multiple data streams~\cite{blikstein2013mla,worsley2018mla,ochoa2016mla}.
Prior work has studied multimodal signals such as gaze, gesture, posture, and speech prosody as indicators of engagement and cognition~\cite{dmello2015multimodal}.
In classroom settings, researchers have combined video and audio to assess teaching quality and detect student affect or confusion~\cite{bosch2016detecting}.
More recently, multimodal deep learning has been applied directly to classroom video.
Hou \emph{et al}.~\cite{hou2025multimodal} proposed a text-centered multi-task model for discourse quality assessment in mathematics but found that visual features did not consistently improve performance over transcript-only baselines.
This mixed evidence may reflect the nature of mathematics instruction, where key information is often explicitly verbalized.
Science classrooms provide a significantly stronger case for multimodal analysis: CIP involve physical artifacts, data representations, and gestural references to shared inscriptions~\cite{latour1986visualization,windschitl2012proposing} that transcripts alone cannot capture.

\Paragraph{Foundation Models for Classroom Analytics}
Large Language Models (LLMs) have been widely studied for educational tasks, presenting both opportunities and challenges~\cite{brown2020language,kasneci2023chatgpt}.
Multimodal LLMs (MLLMs) that jointly process text with images and/or video, including GPT-4V~\cite{openai2023gpt4}, Gemini~\cite{team2023gemini}, and LLaVA~\cite{liu2023llava}, further expand the design space for classroom analytics.
Prompting strategies substantially affect model behavior: few-shot prompting~\cite{brown2020language} provides exemplars, while chain-of-thought prompting~\cite{wei2022chain,kojima2022large} can improve reasoning on complex tasks.
In education, foundation models have been explored for feedback generation, tutoring dialogue, and assessment~\cite{dai2023llmfeedback,macina2023mathdial,tack2023bea}, yet relatively few studies examine MLLMs on authentic classroom video.
Wang \emph{et al}.~\cite{wang2023chatgpt} compared ChatGPT with fine-tuned BERT for talk move classification and found that zero-shot LLMs underperformed supervised models, motivating the need for careful evaluation designs.
Our work addresses these gaps by benchmarking state-of-the-art LLMs and MLLMs for CIP coding in science classroom video, comparing transcript-only versus multimodal inputs under zero-shot, few-shot, and chain-of-thought prompting, and evaluating not only accuracy but also the quality of evidence underlying model predictions.
%
\definecolor{low1}{HTML}{FFF3E0}  
\definecolor{low2}{HTML}{FFE0B2}  
\definecolor{high3}{HTML}{E3F2FD} 
\definecolor{high4}{HTML}{BBDEFB} 
\definecolor{headerblue}{HTML}{E8EAF6} 
\definecolor{practiceCol}{HTML}{FAFAFA} 

\begin{table}[t]
\centering
\caption{Core Instructional Practices (CIP) framework~\cite{windschitl2012proposing}. Pedagogical sophistication increases from Level 1 (surface features) to Level 4 (model-based inquiry). Shading distinguishes the Low (Levels 1--2, \colorbox{low2}{\strut orange}) and High (Levels 3--4, \colorbox{high3}{\strut blue}) categories used for the binary sophistication probe.}
\label{tab:cip_framework}
\scriptsize
\renewcommand{\arraystretch}{1.4}
\setlength{\tabcolsep}{3pt}
\begin{tabular}{>{\columncolor{practiceCol}\centering}m{0.6cm} >{\columncolor{low1}\raggedright\arraybackslash}m{2.6cm} >{\columncolor{low2}\raggedright\arraybackslash}m{2.6cm} >{\columncolor{high3}\raggedright\arraybackslash}m{2.6cm} >{\columncolor{high4}\raggedright\arraybackslash}m{2.6cm}}
\toprule
\rowcolor{headerblue}
\cellcolor{practiceCol}  & \textbf{Level 1} & \textbf{Level 2} & \textbf{Level 3} & \textbf{Level 4} \\
\midrule
\textbf{BI} & 
\textbf{Topics, vocabulary, ``things.''} Students name, label, identify using correct vocabulary. & 
\textbf{Observable process.} Focus on ``what is changing'' or how conditions affect an event. & 
\multicolumn{2}{>{\columncolor{high3}\raggedright\arraybackslash}m{5.4cm}}{\textbf{Explanatory model focus.} Focus on unobservable processes/entities and relationships among concepts. Link to observable phenomena to develop explanatory models.} \\
\addlinespace[4pt]
\textbf{D1} & 
\textbf{Monitor and reteach.} Check for ``correct'' conceptions; one-on-one tutoring or IRE pattern. & 
\textbf{Elicit initial understandings.} Draw out students' hypotheses and questions about scientific ideas. & 
\multicolumn{2}{>{\columncolor{high3}\raggedright\arraybackslash}m{5.4cm}}{\textbf{Adapt to student ideas.} Pose open-ended tasks or puzzling events. Use students' language to shape conversations.} \\
\addlinespace[4pt]
\textbf{D2} & 
\textbf{Focus on procedure.} Describe procedures and experimental setups; downplay concepts. & 
\textbf{Discover/confirm ideas.} ``Proof of concept'' activities; acquire accepted facts and laws. & 
\textbf{Link concepts across investigations.} Seed new concepts; students derive explanatory language. & 
\textbf{Model-based inquiry.} Use evolving models as reference before, during, and after inquiry. \\
\addlinespace[4pt]
\textbf{D3} & 
\textbf{No press for explanation.} No explanation required; ``explain'' means ``justify.'' & 
\textbf{``What happened.''} Describe variables, group differences, trends, or observations. & 
\textbf{``How/partial why.''} Hypothesize and predict system behavior. & 
\textbf{Causal explanation.} Use unobservables to construct causal stories; discuss ``what counts'' as evidence. \\
\bottomrule
\end{tabular}
\end{table}

\begin{figure}[t]
\begin{center}
\includegraphics[width=1\textwidth]{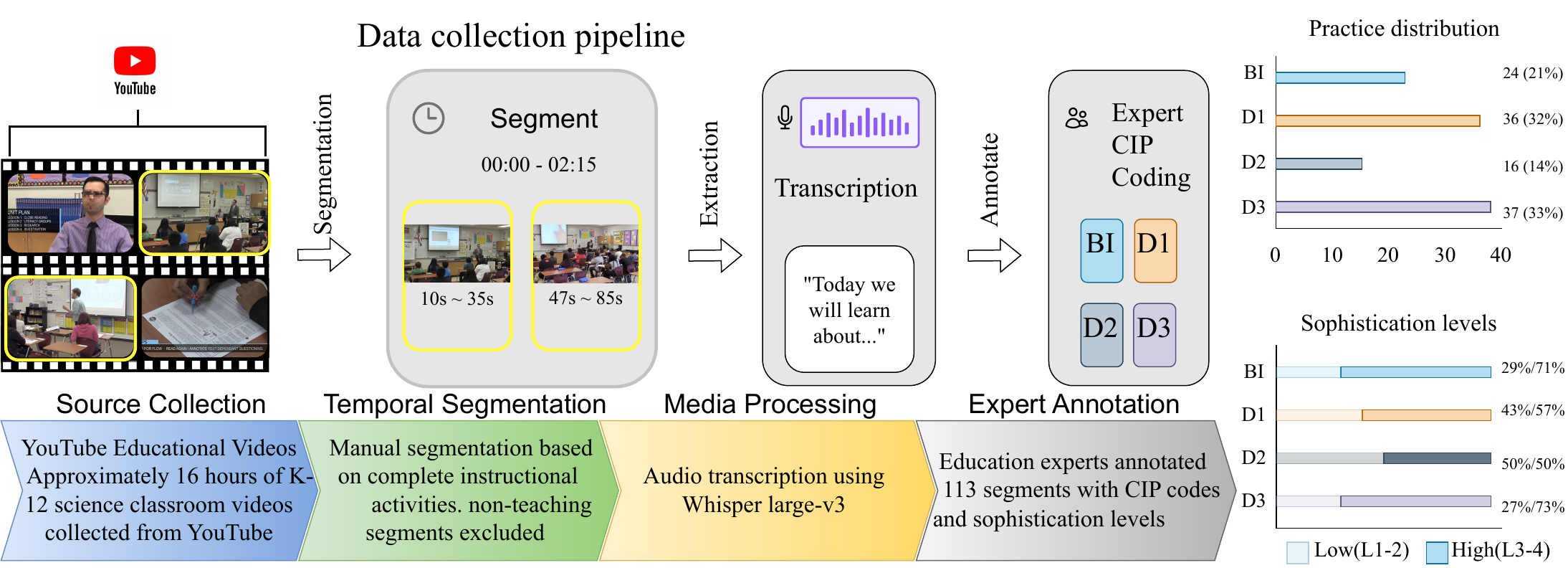}
\end{center}
\vspace{-5mm}
\caption{\textbf{Overview of the SciIBI benchmark construction.} The pipeline involves sourcing NGSS-aligned videos, temporal segmentation based on instructional activities, and consensus-based expert annotation. The final dataset features 113 clips with a naturalistic distribution across four Core Instructional Practices (CIP) and binary sophistication levels (Low vs. High).}
\label{fig2}
\vspace{-3mm}
\end{figure}

\section{Method}
\label{sec:method}

\subsection{Framework and Tasks}
\label{sec:method_task}

\Paragraph{Core Instructional Practices (CIP)}
We operationalize science classroom discourse analysis using the \emph{Core Instructional Practices} framework from Windschitl \emph{et al.}~\cite{windschitl2012proposing}.
The framework specifies four practices that support inquiry-based instruction and provides a performance progression for each, ranging from basic to ambitious enactments.

\Paragraph{Primary Task: 4-way CIP Classification}
Given a classroom clip (with transcript and optional multimodal inputs), a model predicts the dominant core instructional practice:
\begin{itemize}
    \item \textbf{Big Idea (BI)}: selecting and framing big ideas as models that link unobservable processes to phenomena;
    \item \textbf{D1}: eliciting students' initial ideas to adapt instruction;
    \item \textbf{D2}: guiding students to use theories and models to make sense of observations from the inquiry activity;
    \item \textbf{D3}: pressing students for evidence-based explanations that coordinate claims with evidence.
\end{itemize}
Full coding criteria and level descriptors are summarized in Tab.~\ref{tab:cip_framework}.

\Paragraph{Secondary Probe: Within-practice Sophistication}
Beyond 4-way classification, we test whether models distinguish lower- versus higher-sophistication enactments \emph{within} each practice.
We derive a binary label from the Windschitl progression:
for BI and D1 (3 levels each), \emph{Low} = Levels 1--2 and \emph{High} = Level 3;
for D2 and D3 (4 levels each), \emph{Low} = Levels 1--2 and \emph{High} = Levels 3--4.
This probe requires sensitivity to pedagogical intent rather than surface cues.

\Paragraph{Evidence Requirement}
For every prediction, models must provide verifiable supporting evidence:
(i)~\emph{text evidence}: a quoted span or sentence indices from the transcript, and/or
(ii)~\emph{temporal evidence}: a timestamp interval referencing visual cues.
This enables interpretability, human verification, and evidence-quality evaluation (Sec.~\ref{sec:exp_evidence}).

\subsection{The SciIBI Benchmark}
\label{sec:method_data}
We introduce \textbf{SciIBI} (\textbf{Sci}ence \textbf{I}nquiry-\textbf{B}ased \textbf{I}nstruction), a video benchmark for evaluating automated instructional practice coding in K--12 science classrooms.
To our knowledge, SciIBI is the first benchmark targeting science-specific discourse practices with multimodal inputs and evidence-based evaluation.

\Paragraph{Source Videos}
We began by surveying 120 YouTube channels related to K--12 science education, selecting those with professional classroom demonstration videos.
From these, we identified channels containing NGSS-aligned science lessons from public educational institutions (\emph{e.g.}, NSTA~\cite{nsta}, SBCUSD~\cite{sbcusd}), non-profit educational initiatives (\emph{e.g.}, Edutopia~\cite{edutopia}), commercial educational publishers (\emph{e.g.}, Carolina Science~\cite{carolinascience}, NGL Educational Consultants~\cite{ngl}, HMH Education~\cite{hmh}), and individual classroom teachers (\emph{e.g.}, Jerrid Kruse~\cite{jerridkruse}).
Within each channel, we selected NGSS-aligned videos that (i)~focus on authentic teaching demonstrations, (ii)~feature both teacher and student interactions, and (iii)~contain minimal non-classroom audio such as narrator voiceovers, background music, or interview segments.
To ensure diversity, we sampled across grade bands (elementary, middle, high school) and science disciplines (life, physical, earth science).
Quality filtering removed videos with poor audio, excessive background noise, or extended off-task segments, yielding $82$ lessons totaling approximately $16.35$ hours of instruction.

\Paragraph{Segmentation and Transcription}
Each lesson was segmented into several clips containing meaningful instructional segments, based on the occurrence of core instructional practices defined by the CIP analytic framework. The resulting 113 clips range from 6 to 600 seconds (mean: 92s, median: 63s). Transcripts were generated using Whisper (large-v3~\cite{radford2022whisper}) and lightly corrected for speaker attribution and domain-specific terminology; timestamps were preserved to support evidence localization.

\Paragraph{Annotation}
All clips were independently coded by two researchers---a doctoral student and a faculty member---both with professional training in the CIP framework.
All clips were coded to consensus following a predefined rubric.
Annotators jointly reviewed clips and resolved differences through discussion to arrive at a single agreed-upon label for both the 4-way CIP classification (BI/D1/D2/D3) and the binary sophistication probe.
Because coding was performed to consensus rather than fully independent double-coding, we do not report inter-rater reliability statistics; this consensus-based procedure is standard in theory-driven classroom coding when constructs are conceptually overlapping and require interpretive judgment~\cite{windschitl2012proposing}.
For instance, a clip showing a teacher guiding students to review particle properties while revising atomic models was initially split between BI and D1; after discussion, it was coded as D1-Level~3, as the primary function of this discourse was eliciting student ideas and adapting instruction.

\Paragraph{Statistics}
The final benchmark comprises 113 clips totaling approximately 3 hours of classroom video.
The label distribution reflects naturalistic variation: BI ($n{=}24$, 21\%), D1 ($n{=}36$, 32\%), D2 ($n{=}16$, 14\%), and D3 ($n{=}37$, 33\%).
Fig.~\ref{fig2} illustrates the data pipeline and provides detailed statistics.

\subsection{Models and Input Conditions}
\label{sec:method_models}

We evaluate a diverse set of large language models spanning text-only LLMs and multimodal LLMs (MLLMs) to assess how model architecture and input modality affect instructional practice coding.
Our evaluation includes both open-source models deployed locally (Mistral-7B~\cite{jiang2023mistral7b}, Llama-3.3-70B~\cite{grattafiori2024llama3herdmodels}, 
GPT-OSS-20B~\cite{openai2025gptoss120bgptoss20bmodel}, InternVL3-78B~\cite{zhu2025internvl3exploringadvancedtraining}) and proprietary models accessed via API (GPT-4o~\cite{openai2024gpt4ocard}, Claude Sonnet 4.5~\cite{anthropic2025claude}, Gemini-2.5-Pro~\cite{comanici2025gemini25pushingfrontier}, Qwen3-VL-235B~\cite{bai2025qwen3vltechnicalreport}).
This selection covers a range of model scales and multimodal capabilities, enabling comparison across the current landscape of foundation models.
Full model identifiers, versions, and inference settings are reported in Sec.~\ref{sec:exp_setup}.

To isolate the contribution of each modality, we evaluate models under three input conditions.
\textbf{Text-only (T)} provides the ASR-generated transcript, testing whether linguistic content alone suffices for practice classification.
\textbf{Vision + Text (V+T)} adds uniformly sampled video frames, enabling models to observe classroom artifacts, gestures, and spatial arrangements that may disambiguate practice categories.
This design allows us to quantify whether---and for which practices---visual signals improve classification beyond text.

\subsection{Prompting Protocol}
\label{sec:method_prompting}

We evaluate three prompting strategies that reflect realistic deployment scenarios where task-specific fine-tuning is infeasible: \textbf{zero-shot}, \textbf{few-shot}, and \textbf{chain-of-thought (CoT)}~\cite{brown2020language,wei2022chain}.
All strategies use identical CIP category definitions and require structured outputs to ensure fair comparison.
In the \textbf{zero-shot} setting, models receive only task instructions and category definitions, testing their ability to apply pedagogical concepts from pre-training alone.
The \textbf{few-shot} setting augments the prompt with four exemplars (one per practice category), held constant across models for controlled comparison; exemplars were selected from pilot misclassifications to maximize informativeness.
The \textbf{CoT} setting instructs models to reason step-by-step before producing a classification; however, we evaluate only the final label and evidence fields, avoiding inflated performance from unverifiable reasoning traces.
All prompts require a structured JSON response containing \texttt{category}, \texttt{sophistication}, and \texttt{evidence} (transcript spans and/or timestamps).
Outputs failing schema validation receive one retry with a format reminder; responses that remain unparseable are logged as failures and excluded from accuracy calculations but reported separately to characterize model reliability.

\section{Experiments}
\label{sec:experiments}

\subsection{Experimental Setup}
\label{sec:exp_setup}

\Paragraph{Implementation}
Experiments were conducted on 6 NVIDIA L40S GPUs.
Open-source models were deployed locally; we applied 4-bit quantization via BitsAndBytes for models exceeding 40B parameters.
All models used deterministic decoding (\texttt{temperature=0.0}) with \texttt{max\_new\_tokens=1024}.

\Paragraph{Models}
We evaluate eight models spanning a range of scales and modality support.
\emph{Open-source models} include Mistral-7B-Instruct-v0.3~\cite{jiang2023mistral7b}, Llama-3.3-70B-Instruct~\cite{grattafiori2024llama3herdmodels}, GPT-OSS-20B~\cite{openai2025gptoss120bgptoss20bmodel}, and InternVL3-78B~\cite{zhu2025internvl3exploringadvancedtraining}, the latter being the only open-source MLLM in our study.
\emph{Proprietary models} accessed via API include GPT-4o~\cite{openai2024gpt4ocard}, Claude Sonnet 4.5~\cite{anthropic2025claude}, Gemini-2.5-Pro~\cite{comanici2025gemini25pushingfrontier}, and Qwen3-VL-235B-A22B~\cite{bai2025qwen3vltechnicalreport}.
This selection enables comparison across model scales (7B--235B parameters), architectures (text-only LLMs vs.\ MLLMs), and access paradigms (local deployment vs.\ API).

\Paragraph{Multimodal Inputs}
For the V+T condition, we uniformly sample 1 frame per second, capped at 30 frames for longer clips, and resize frames to fit model input constraints.

\Paragraph{Metrics and Invalid Outputs}
We report classification accuracy as the primary metric, computed as the percentage of clips where the predicted label matches any valid ground truth label (accounting for clips with multiple acceptable codings). We also report macro-averaged F1, precision, and recall scores for completeness.

\subsection{Diagnostic Findings: Classification Accuracy and Error Patterns}

Tab.~\ref{tab:text_results} presents classification performance using transcript-only input across all models and prompting strategies. We first establish baseline accuracy to contextualize subsequent failure analysis. Zero-shot accuracies range from 39.1\% to 46.4\%, indicating that current models lack reliable pedagogical reasoning for CIP coding. This gap motivates our deeper investigation into why models fail rather than simply quantifying how much they fail.

\begin{table}[t]
\centering
\caption{Classification performance on the \textbf{SciIBI} benchmark. We report Accuracy (Acc) and Macro-F1 scores (\%) across eight models using Text-only (T) and Text+Vision (TV) inputs under Zero-shot, Few-shot, and CoT prompting. 
}
\label{tab:text_results}
\small
\setlength{\tabcolsep}{3.5pt}
\renewcommand{\arraystretch}{1.0}
\begin{tabular}{l r c cc cc cc}
\toprule
& & & \multicolumn{2}{c}{Zero-shot} & \multicolumn{2}{c}{Few-shot} & \multicolumn{2}{c}{CoT} \\
\cmidrule(lr){4-5} \cmidrule(lr){6-7} \cmidrule(lr){8-9}
\multirow{-2}{*}{Model} & \multirow{-2}{*}{Size} & \multirow{-2}{*}{Mod} & Acc$\uparrow$ & F1$\uparrow$& Acc$\uparrow$& F1$\uparrow$& Acc$\uparrow$ & F1$\uparrow$\\
\midrule
\rowcolor{gray!5}
\multicolumn{9}{l}{\textit{Proprietary (API)}} \\
GPT-4o~\cite{openai2024gpt4ocard} & — & T & 45.4 & 43.6 & 45.5 & 42.9 & 50.9 & 48.4 \\
GPT-4o~\cite{openai2024gpt4ocard} & — & TV & 49.1 & 46.1 & 47.7 & 46.2 & \textbf{51.9} & \textbf{48.5} \\
Claude Sonnet 4.5~\cite{anthropic2025claude} & — & T & 43.8 & 41.2 & 45.5 & 43.8 & 46.4 & 44.4 \\
Gemini-2.5-Pro~\cite{comanici2025gemini25pushingfrontier} & — & T & 39.3 & 38.8 & 42.0 & 40.3 & 42.9 & 41.3 \\
Gemini-2.5-Pro~\cite{comanici2025gemini25pushingfrontier} & — & TV & 42.9 & 39.9 & 42.9 & 40.3 & 48.2 & 44.6 \\
Qwen3-VL-235B~\cite{bai2025qwen3vltechnicalreport} & 235B & T & 46.4 & 40.8 & 48.2 & 43.5 & 45.5 & 41.7 \\
Qwen3-VL-235B~\cite{bai2025qwen3vltechnicalreport} & 235B & TV & 45.5 & 42.3 & 47.3 & 43.2 & 49.1 & 43.3 \\
\midrule
\rowcolor{gray!5}
\multicolumn{9}{l}{\textit{Open-source (Local)}} \\
Mistral-7B~\cite{jiang2023mistral7b} & 7B & T & 40.2 & 33.2 & 36.6 & 31.3 & 34.8 & 30.3 \\
GPT-OSS-20B~\cite{openai2025gptoss120bgptoss20bmodel} & 20B & T & 39.1 & 37.4 & 37.7 & 35.5 & 36.6 & 35.6 \\
Llama-3.3-70B~\cite{grattafiori2024llama3herdmodels} & 70B & T & 44.6 & 42.5 & 47.3 & 44.8 & 48.2 & 45.4 \\
InternVL3-78B~\cite{zhu2025internvl3exploringadvancedtraining} & 78B & T & 46.4 & 43.6 & 47.3 & 44.0 & 47.3 & 45.3 \\
InternVL3-78B~\cite{zhu2025internvl3exploringadvancedtraining} & 78B & TV & \textbf{53.6} & \textbf{47.8} & \textbf{50.9} & \textbf{45.3} & 50.9 & 46.2 \\
\bottomrule
\end{tabular}
\end{table}

\Paragraph{Overall Performance}
Zero-shot accuracies range from 39.1\% (GPT-OSS) to 53.6\% (InternVL3), substantially below the ${\sim}$79\% F1 reported for mathematics discourse coding on TalkMoves~\cite{suresh2022fine}.
This gap persists even with strong general-purpose models, suggesting that CIP coding requires instructional-function judgments that are not captured by surface lexical cues---for instance, distinguishing whether a teacher's question is eliciting prior ideas (D1) versus pressing for causal explanation (D3).

\Paragraph{Prompting Effects}
Chain-of-thought prompting benefits larger models: GPT-4o gains +5.5pp (45.4\%$\rightarrow$50.9\%), Llama-3.3-70B gains +3.6pp, and Claude gains +2.7pp.
However, CoT \emph{degrades} smaller models: Mistral-7B drops $-$5.4pp and GPT-OSS drops $-$2.4pp.
This asymmetry suggests that step-by-step reasoning helps when models have sufficient capacity for pedagogical judgment, but introduces errors when capacity is limited.
Few-shot prompting yields modest, inconsistent gains (+1--3pp), indicating that four exemplars provide limited signal for nuanced instructional distinctions.

\Paragraph{Sophistication Probe}
For the secondary task of predicting sophistication levels (1--4 within each practice), we evaluated InternVL3-78B predictions against ground truth.
The model achieved 39.1\% accuracy on the 4-level task and 69.1\% on the binary Low/High probe (Tab.~\ref{tab:sophistication}).
Predictions skewed toward higher sophistication: Level~3 (42.2\%) and Level~4 (22.0\%) were predicted most frequently, while Level~1 appeared in only 9.2\% of outputs.
This suggests models may default to ``optimistic'' assessments, possibly because high-sophistication discourse is more explicitly marked by causal language and evidence coordination.


\begin{table}[t]
\centering
\caption{Sophistication level prediction (InternVL3-78B, Vision+Text). Accuracy is reported for both fine-grained (4-Level) and binary (Low/High) tasks. The "Require L1" condition highlights the model's struggle to identify lower-level practices, often defaulting to optimistic higher-level predictions.}
\label{tab:sophistication}
\small
\setlength{\tabcolsep}{6pt}         
\renewcommand{\arraystretch}{1}  

\begin{tabular}{
  >{\raggedright\arraybackslash}p{0.16\linewidth}
  >{\raggedright\arraybackslash}p{0.14\linewidth}
  c c c c
}
\toprule
\textbf{Condition} & \textbf{Metric} & \textbf{BI} & \textbf{D1} & \textbf{D2} & \textbf{D3} \\
\midrule
\multirow{2}{*}{Ignore L1}  & 4-Level & 50.0 & 28.6 & 50.0 & 37.8 \\
                           & Binary  & 66.7 & 54.3 & 85.7 & 78.4 \\
\midrule
\multirow{2}{*}{Require L1} & 4-Level & 45.5 & 34.8 & 50.0 & 33.3 \\
                            & Binary  & 45.5 & 52.2 & 75.0 & 85.7 \\
\midrule
\multirow{2}{*}{$N$}       & Ignore L1  & 24 & 35 & 14 & 37 \\
                           & Require L1 & 11 & 23 & 4  & 21 \\
\bottomrule
\end{tabular}
\end{table}

\subsection{Modality Ablation}

Tab.~\ref{tab:vision_results} compares transcript-only versus vision+text inputs for MLLMs.

\begin{table}[t]
\centering
\caption{ Impact of visual modality on classification accuracy. Comparison of MLLMs using Vision+Text inputs versus their text-only baselines. $\Delta$ denotes the percentage point change relative to the text-only zero-shot performance.}
\label{tab:vision_results}
\small
\setlength{\tabcolsep}{5pt}
\begin{tabular}{l cc cc cc}
\toprule
& \multicolumn{2}{c}{Zero-shot} & \multicolumn{2}{c}{Few-shot} & \multicolumn{2}{c}{CoT} \\
\cmidrule(lr){2-3} \cmidrule(lr){4-5} \cmidrule(lr){6-7}
\multirow{-2}{*}{Model} & Acc & $\Delta$ & Acc & $\Delta$ & Acc & $\Delta$ \\
\midrule
InternVL3-78B~\cite{zhu2025internvl3exploringadvancedtraining} & \textbf{53.6} & +7.1 & 50.9 & +4.5 & 50.9 & +4.5 \\
GPT-4o~\cite{openai2024gpt4ocard} & 49.1 & +3.7 & 47.7 & +2.3 & 51.9 & +6.5 \\
Gemini-2.5-Pro~\cite{comanici2025gemini25pushingfrontier} & 42.9 & +3.6 & 42.9 & +3.6 & 48.2 & +\textbf{8.9} \\
Qwen3-VL-235B~\cite{bai2025qwen3vltechnicalreport} & 45.5 & $-$0.9 & 47.3 & $+$0.9 & 49.1 & +2.7 \\
\bottomrule
\end{tabular}
\end{table}

\Paragraph{Best Result}
The highest accuracy is \textbf{53.6\%}, achieved by InternVL3-78B with vision+text in zero-shot mode---a +7.1pp gain over transcript-only.
The largest gains appear in clips where the transcript omits the referenced representation (\emph{e.g.}, whiteboard diagram, data table), consistent with artifact-mediated sensemaking that transcripts cannot capture.

\Paragraph{Model Variation}
The benefit of visual input varies substantially: InternVL3 gains +7.1pp, GPT-4o gains +3.7pp, and Gemini gains +3.6pp, but Qwen3-VL \emph{decreases} by $-$0.9pp in zero-shot mode.
This inconsistency indicates that \emph{not all MLLMs effectively integrate visual information for pedagogical reasoning; architecture and training data likely influence whether video frames help or introduce noise}.

\Paragraph{Per-category Analysis}
The best configuration (InternVL3-78B, V+T, zero-shot) shows substantial variation across practice categories.
D1 (eliciting student ideas) achieves the highest accuracy at 57.1\%, likely because verbal elicitation patterns---open-ended questions, invitations to share thinking---are relatively distinctive in transcripts.
D3 (pressing for explanation) follows at 35.1\%, while BI (big idea framing) reaches 29.2\%.
D2 (sensemaking around activity) is the most challenging at only 18.8\%, consistent with our hypothesis that this practice depends heavily on artifact-mediated interaction requiring visual grounding.
The large gap between D1 and D2 underscores the limits of current MLLMs: even with video input, models struggle to interpret the pedagogical function of physical materials and student manipulations.

\subsection{Evidence Quality Analysis}
\label{sec:exp_evidence}

Beyond classification accuracy, we assess whether models provide appropriate justification for their predictions---a requirement for trustworthy deployment in teacher-facing tools.

\Paragraph{EQS Framework}
We designed an Evidence Quality Score (EQS) with three dimensions, each rated on a 1--3 scale:
\emph{Alignment}---does the cited evidence support the predicted category?
\emph{Sufficiency}---is enough evidence provided to justify the prediction?
\emph{Specificity}---does the evidence reference concrete transcript spans or timestamps rather than vague descriptions?
Two raters independently scored a stratified sample of 60 model outputs, balanced across practice categories and prediction correctness (15 correct + 15 incorrect per top-2 model).
Inter-rater agreement was substantial to near-perfect across dimensions ($\kappa_{\text{grounding}}=0.91$, $\kappa_{\text{specificity}}=0.92$, $\kappa_{\text{alignment}}=0.73$), supporting the reliability of the rubric while reflecting the interpretive complexity of aligning model explanations with theoretical CIP definitions; disagreements were resolved by discussion.
Tab.~\ref{tab:eqs} reports the averaged Evidence Quality Scores for the two top-performing models across the three rubric dimensions, along with the overall mean EQS.


\begin{table}[t]
\centering
\caption{Evidence Quality Scores (EQS). Human-rated scores (1--3 scale) assessing whether the model's cited evidence is Aligned with the label, Sufficient to justify it, and Specific (referencing timestamps/quotes). Results indicate that higher classification accuracy does not always correlate with higher quality reasoning.}
\label{tab:eqs}
\small
\setlength{\tabcolsep}{7pt}         
\renewcommand{\arraystretch}{1}  

\begin{tabular}{
  >{\raggedright\arraybackslash}p{0.24\linewidth}
  c c c c
}
\toprule
Model & Align. & Suff. & Spec. & Mean EQS \\
\midrule
InternVL3-78B~\cite{zhu2025internvl3exploringadvancedtraining} & 2.53 & 2.02 & 2.65 & 2.40 \\
GPT-4o~\cite{openai2024gpt4ocard} & 2.67 & 2.69 & 2.64 & 2.67 \\
\bottomrule
\end{tabular}
\end{table}

\Paragraph{Accuracy vs.\ Evidence Quality}
EQS reveals patterns that accuracy alone cannot detect:
\emph{High-accuracy, low-EQS (right for wrong reasons).}
Some correct predictions cite irrelevant or superficial evidence.
For example, a clip correctly labeled D3 was justified by ``the teacher asks students to explain,'' without referencing the specific press for evidence and reasoning that distinguishes D3 from D1.
Such cases suggest reliance on lexical shortcuts rather than genuine pedagogical understanding.
\emph{Low-accuracy, high-EQS (reasonable mistakes).}
Some incorrect predictions cite appropriate evidence but arrive at the wrong label due to framework ambiguity.
A clip coded as D1 was predicted as D3 with evidence pointing to ``teacher questions probing student reasoning''---a defensible interpretation given overlapping surface features.

EQS thus exposes ``right answer, wrong evidence'' cases that accuracy metrics miss, which matters for teacher-facing trust: a model that succeeds through shortcuts may fail unpredictably on novel clips.

\subsection{Failure Analysis and Deployment Guidance}
\begin{figure}[t]
\centering
\includegraphics[width=1\textwidth]{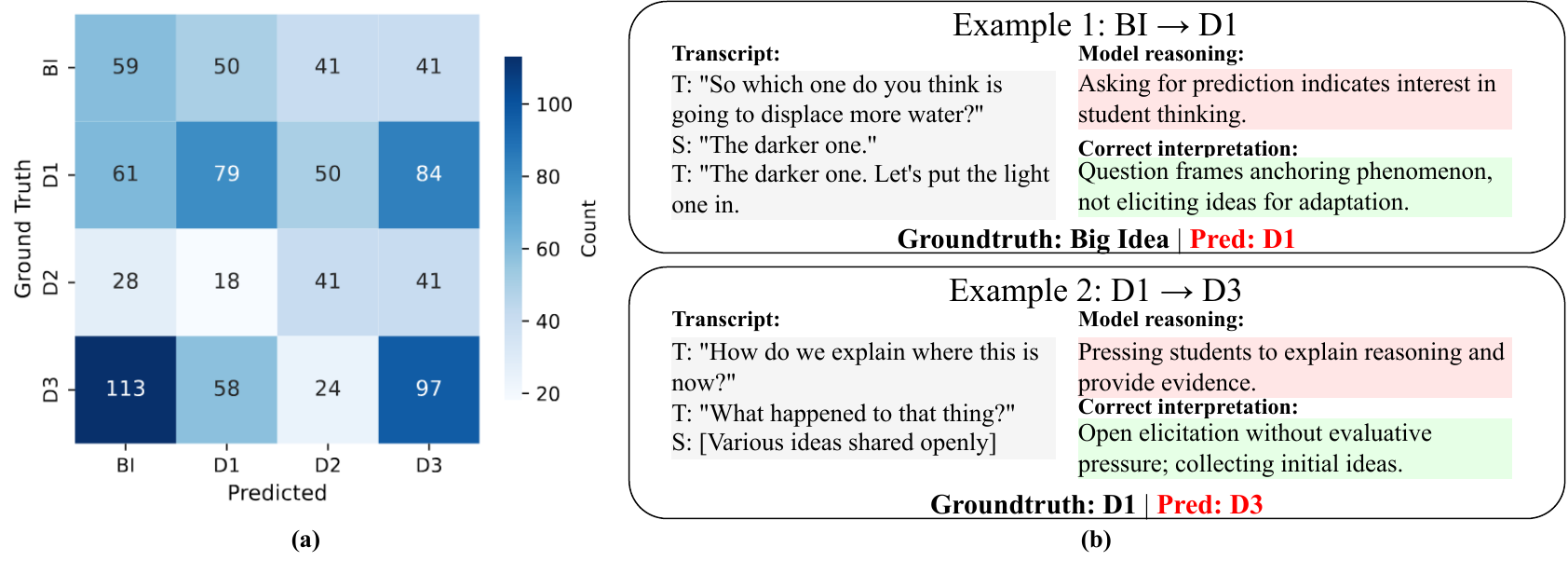}
\vspace{-5mm}
\caption{Failure analysis of text-only models. (a) Aggregated confusion matrix reveals systematic confusion between eliciting ideas (D1) and pressing for explanation (D3). (b) Representative errors illustrate how models often rely on surface keywords (\emph{e.g.}, "asking for prediction") rather than the underlying pedagogical function, leading to misinterpretations of instructional intent.}
\vspace{-3mm}
\label{fig:confusion}
\end{figure}

\Paragraph{Confusion Patterns}
Fig.~\ref{fig:confusion} (a) visualizes the aggregated confusion matrix across all models (zero-shot, text-only).
Two dominant error patterns emerge: D3$\rightarrow$BI (113 cases) and D1$\rightarrow$D3 (84 cases).
Both confusions share a root cause: models match surface linguistic features---explanation-related vocabulary for D3/BI, question forms for D1/D3---without distinguishing the underlying pedagogical function.
This suggests current LLMs lack the instructional reasoning needed to separate \emph{what is said} from \emph{why it is said}.

\Paragraph{Illustrative Failures}
To contextualize these statistical patterns, Fig.~\ref{fig:confusion} (b) presents two representative errors from InternVL3-78B to illustrate common failure modes regarding the misclassification of teacher questions and explanations:

\emph{Example 1: BI misclassified as D1.}
In a clip where the teacher asks ``What do you think will happen when we drop both objects?'', the model reasoned: ``The teacher is asking for student predictions, indicating interest in understanding student thinking.''
However, this question serves to frame the anchoring phenomenon (BI), not to surface prior conceptions for instructional adaptation (D1).
The model correctly identified the question form but missed the instructional purpose.

\emph{Example 2: D1 misclassified as D3.}
In a first-grade light investigation, the model stated: ``The teacher is pressing students to explain their reasoning and provide evidence.''
In fact, the teacher was only collecting students' ideas about the evidence and did not require them to provide a complete explanation based on the evidence.
\section{Limitations and Future Work}
\label{sec:limitations}

Our benchmark, while enabling reproducible evaluation, relies on publicly available YouTube videos that may not fully capture the unscripted variability of everyday classroom instruction; future work should prioritize in-situ recordings collected in partnership with schools under authentic conditions.
The substantial performance gap relative to mathematics benchmarks suggests that zero-shot prompting alone is insufficient, and parameter-efficient fine-tuning via LoRA~\cite{hu2021loralowrankadaptationlarge} or adapters~\cite{houlsby2019parameterefficienttransferlearningnlp} may be necessary to capture the subtle decision boundaries between CIP categories.
While current accuracy levels preclude fully autonomous classification, models can meaningfully accelerate human workflows by retrieving candidate evidence spans and supporting confidence-based prioritization.
We envision an \emph{evidence-first} interface where model outputs serve as searchable annotations rather than authoritative labels, positioning AI as a retrieval assistant that augments expert judgment rather than replacing it.

\section{Conclusion}

In this work, we introduce \textbf{SciIBI}, the first video benchmark for science classroom CIP coding, comprising 113 clips annotated with Core Instructional Practices and sophistication levels by trained education researchers.
We benchmark eight state-of-the-art LLMs and MLLMs across input modalities and prompting strategies, demonstrating that CIP coding exposes fundamental limitations in current models' ability to reason about instructional purpose beyond surface linguistic patterns.
Three key insights emerge from our evaluation: visual input yields modest gains concentrated in artifact-mediated clips, though not all MLLMs effectively leverage video; chain-of-thought prompting benefits large models while degrading smaller ones; and evidence-based evaluation reveals critical cases where models succeed through lexical shortcuts rather than genuine pedagogical understanding.
These findings establish science classroom analysis as a challenging frontier for multimodal AI and suggest that human-AI collaboration through evidence-first interfaces offers the most viable near-term path for deployment.

\bibliographystyle{splncs04}
\bibliography{ref}

\end{document}